\documentstyle[amsfonts,preprint,aps]{revtex}

\begin{document}
\title{{\LARGE The diffusion equation and the principle of minimum Fisher
information}}
\author{Marcel Reginatto}
\address{Environmental Measurement Laboratory, U.S. Department of Energy, \\
201 Varick Street, 5th Floor, New York, New York 10014-4811\\
(phone: (212) 620-3634, fax: (212) 620-3600, e-mail: mreg@eml.doe.gov)}
\author{Florian Lengyel}
\address{Department of Mathematics, The Graduate School and University Center of CUNY,%
\\
33 West 42nd Street, New York, NY 10036}
\date{\today}
\maketitle

\begin{abstract}
It is shown that the diffusion equation and its adjoint (time reversed)
equation can be derived with only a few assumptions, using an
information-theoretic approach based on the principle of minimum Fisher
information.


\end{abstract}

\section{Introduction}

The derivation of the diffusion equation from a fixed end-point variational
principle is well known \cite{Morse and Feshbach}. \ The Lagrangian that is
normally used leads simultaneously to two equations for two real functions:
the diffusion equation for a function $\psi $, and its adjoint (time
reversed) equation for a function $\psi ^{\ast }$.\ This Lagrangian is
usually introduced formally, without physical justification (consider the
following quote from Ref. \cite{Morse and Feshbach}: ``The introduction of
the mirror-image field $\psi ^{\ast }$, in order to set up a Lagrange
function from which to obtain the diffusion equation, is probably too
artificial a procedure to expect to obtain much of physical significance
from it''). \ We wish to show that this Lagrangian results from applying an
information-theoretic approach to the solution of the following
interpolation problem.

Consider an experiment, where the probability density $\rho (x,t)$ and the
average velocity field $v(x,t)$ of a cloud of particles of mass $m$ is
measured at times $t_{0}$ and $t_{1}$(for simplicity, we consider only
motion in one dimension). \ Assume that $\rho $ satisfies the continuity
equation 
\begin{equation}
\frac{\partial \rho }{\partial t}+\frac{\partial }{\partial x}\left( \rho
v\right) =0.  \label{cont}
\end{equation}
Without additional assumptions regarding the dynamics of the system, the
problem of determining the probability density and velocity field at times $t
$ (where $t_{0}<t<$ $t_{1}$) can not be solved, since there are an infinite
number of probability densities and velocity fields that will interpolate
between the values measured at times $t_{0}$ and $t_{1}$. \ However, we
would still like to find {\it best estimates }of $\rho $ and $v$, perhaps by
adding some assumptions about the physical processes that determine the
motion of the cloud of particles, and by using some principle of inference
to select the most likely probability distribution that might describe its
evolution. \ The main result of this paper is to show that the dynamics of
such a system will be determined uniquely by the diffusion equation and its
adjoint equation, 
\begin{equation}
\frac{\partial \psi }{\partial t}=\frac{D}{2m}\frac{\partial ^{2}\psi }{%
\partial x^{2}},  \label{diff}
\end{equation}
\begin{equation}
-\frac{\partial \psi ^{\ast }}{\partial t}=\frac{D}{2m}\frac{\partial
^{2}\psi ^{\ast }}{\partial x^{2}},  \label{adjdiff}
\end{equation}
(where $\psi $ and $\psi ^{\ast }$, defined by eqs. (\ref{phi}) and (\ref
{phistar}), are real functions of $\rho $ and $\sigma $, and $D/2m$ is the
diffusion constant) provided we make the following two assumptions about the
system: that the velocity field can be derived from a potential function $%
\sigma (x,t)$, according to 
\begin{equation}
v=\frac{1}{m}\frac{\partial \sigma }{\partial x},  \label{sigma}
\end{equation}
and that the probability density $\rho $ that interpolates between times $%
t_{0}$ and $t_{1}$ is the one that minimizes the Fisher information $I$
associated with $\rho $, which we define by (see Appendix A) 
\begin{equation}
I=\frac{1}{m}\int_{t_{0}}^{t_{1}}\int_{-\infty }^{+\infty }\frac{1}{\rho }%
\left( \frac{\partial \rho }{\partial x}\right) ^{2}dxdt.  \label{FI}
\end{equation}
The first assumption is equivalent to introducing a particular physical
model, in which the motion of the cloud of particles corresponds to that of
a fluid with no vorticity. \ The second assumption is an
information-theoretical assumption.

\section{Derivation of the diffusion equation from a variational principle}

Eqs. (\ref{cont})\ and (\ref{sigma})\ lead to the continuity equation 
\begin{equation}
\frac{\partial \rho }{\partial t}+\frac{\partial }{\partial x}\left( \rho 
\frac{1}{m}\frac{\partial \sigma }{\partial x}\right) =0.  \label{contsigma}
\end{equation}
Eq. (\ref{contsigma}) can be derived from the Lagrangian $L_{CL}$\ by fixed
end-point variation with respect to $\sigma $, 
\begin{equation}
L_{CL}=-\int_{t_{0}}^{t_{1}}\int_{-\infty }^{+\infty }\rho \left( \frac{%
\partial \sigma }{\partial t}+\frac{1}{2m}\left( \frac{\partial \sigma }{%
\partial x}\right) ^{2}\right) dxdt.  \label{LCL}
\end{equation}
Note also that fixed end-point variation with respect to $\rho $\ leads
trivially to the Hamilton-Jacobi equation, 
\begin{equation}
\frac{\partial \sigma }{\partial t}+\frac{1}{2m}\left( \frac{\partial \sigma 
}{\partial x}\right) ^{2}=0.  \label{H-J}
\end{equation}
Therefore, variation of $L_{CL}$\ with respect to both $\rho $ and $\sigma $
leads to the equations of motion for a classical ensemble, eqs. (\ref
{contsigma}) and (\ref{H-J}). \ There is still considerable freedom in the
choice of probability density that can be used to describe the system, since
it is only subject to eq. (\ref{contsigma}). \ To derive the diffusion
equation and its adjoint, we need to restrict the choice of probability
densities using the principle of minimum Fisher information. \ We consider
therefore the Lagrangian $L_{D}$, 
\begin{equation}
L_{D}=-\int_{t_{0}}^{t_{1}}\int_{-\infty }^{+\infty }\rho \left\{ \frac{%
\partial \sigma }{\partial t}+\frac{1}{2m}\left[ \left( \frac{\partial
\sigma }{\partial x}\right) ^{2}-\frac{D^{2}}{4}\frac{1}{\rho ^{2}}\left( 
\frac{\partial \rho }{\partial x}\right) ^{2}\right] \right\} dxdt.
\label{LD}
\end{equation}
The Lagrangian $L_{D}$ equals $L_{CL}$ plus an additional term proportional
to the Fisher information $I$, 
\begin{equation}
L_{D}=L_{CL}+\frac{1}{2}\frac{D^{2}}{4}I.  \label{LDisLCplusI}
\end{equation}
Fixed end point variation of $L_{D}$ with respect to $\sigma $\ leads once
more to eq. (\ref{contsigma}), while variation with respect to $\rho $\
leads to a modified Hamilton-Jacobi equation that includes a term $Q$ which
is of the form of Bohm's quantum potential \cite{Bohm} (but notice that it
appears here within the context of a classical theory), 
\begin{equation}
\frac{\partial \sigma }{\partial t}+\frac{1}{2m}\left( \frac{\partial \sigma 
}{\partial x}\right) ^{2}+Q=0  \label{modH-J}
\end{equation}
with 
\begin{equation}
Q=-\frac{D^{2}}{4}\left[ \frac{1}{\rho ^{2}}\left( \frac{\partial \rho }{%
\partial x}\right) ^{2}-\frac{2}{\rho }\left( \frac{\partial ^{2}\rho }{%
\partial x^{2}}\right) \right] .  \label{Q}
\end{equation}

Eqs. (\ref{contsigma})\ and (\ref{modH-J}) are identical to eqs. (\ref{diff}%
) and (\ref{adjdiff}) provided we set 
\begin{equation}
\psi =\sqrt{\rho }e^{+\sigma /D},  \label{phi}
\end{equation}
\begin{equation}
\psi ^{\ast }=\sqrt{\rho }e^{-\sigma /D}.  \label{phistar}
\end{equation}
It can be shown (see Appendix B) that the Fisher information $I$\ increases
when $\rho $ is varied while $\sigma $ is kept fixed. \ Therefore, the
solution derived here is the one that minimizes the Fisher information for a
given $\sigma $.

\section{Connection to Brownian motion}

Although $\psi $\ is a solution of the diffusion equation, it will not
correspond in general to the case of Brownian motion. \ Here, $\psi =\sqrt{%
\rho }e^{+\sigma /D}$\ is proportional to the square root of a probability
distribution, while in Brownian motion the probability distribution $\rho $
is the function that satisfies the diffusion equation. \ The $\psi $\ that
we have derived here is essentially the ``wave function'' of Euclidean
quantum mechanics \cite{Zambrini}.

The case of Brownian motion corresponds to a particular solution of eqs. (%
\ref{contsigma})\ and (\ref{modH-J}), one for which 
\begin{equation}
\sigma _{BM}=\frac{D}{2}\ln \rho .  \label{osm}
\end{equation}
In this case, the velocity field then takes the form 
\begin{equation}
v_{BM}=\frac{D}{2m}\frac{\partial \ln \rho }{\partial x}.  \label{osmeq}
\end{equation}
Eq. (\ref{osmeq})\ is known as the osmotic equation, and $v_{BM}$\ is the
osmotic velocity. \ If we substitute $\sigma _{BM}$\ into eqs. (\ref{phi})
and (\ref{phistar}), the ``wave functions'' become 
\begin{equation}
\psi _{BM}=\rho
\end{equation}
\begin{equation}
\psi _{BM}^{\ast }=1
\end{equation}
which solve eqs. (\ref{diff}) and (\ref{adjdiff}) provided the probability
density $\rho $\ is a solution of the diffusion equation. \ One can also
check that eqs. (\ref{contsigma})\ and (\ref{modH-J}) both reduce to 
\begin{equation}
\frac{\partial \rho }{\partial t}+\frac{D}{2m}\frac{\partial ^{2}\rho }{%
\partial x^{2}}=0
\end{equation}
when eq. (\ref{osm}) holds.

\section{Discussion}

It has been shown that the diffusion equation and its adjoint (time
reversed) equation can be derived using an information-theoretic approach
that is based on the principle of minimum Fisher information. \ In the
information-theoretic approach followed here, the emphasis is on using the
principle of minimum Fisher information to complement a physical picture
derived from a particular hydrodynamical model of the system. \ Variation of
the Lagrangian (\ref{LD}) can be interpreted as the minimization of the
Fisher information subject to the constraint that the probability density
satisfy the continuity equation (\ref{contsigma}), which arises naturally in
the hydrodynamical model. \ An alternative approach to the diffusion
equation that also uses minimum Fisher information can be found in Ref. \cite
{Frieden}. \ This derivation, however, differs from the present one in two
crucial respects; in particular, the equation is not derived from a
Lagrangian, and the derivation does not make reference to the hydrodynamical
model.

The approach followed here provides a physically well motivated derivation
of the diffusion equation which distinguishes between physical and
information-theoretical assumptions. \ A similar approach leads to the
Schrodinger \cite{Reginatto} and Pauli \cite{Reginatto2} equations.

\section{Appendix A}

Let $\mu $ be a measure defined on ${\Bbb R}^{n}$, let $P(y^{i})$ be a
probability density with respect to $\mu $ which is a function of $n$
continuous parameters $y^{i}$, and let $P(y^{i}+\Delta y^{i})$ be the
density that results from a small change in the $y^{i}$. Expand the $%
P(y^{i}+\Delta y^{i})$ in a Taylor series, and calculate the cross-entropy
up to the first non-vanishing term, 
\begin{equation}
\int P(y^{i}+\Delta y^{i})\ln \frac{P(y^{i}+\Delta y^{i})}{P(y^{i})}d\mu
(y^{i})\simeq \frac{1}{2}\sum_{j,k}^{n}\left[ \int \frac{1}{P(y^{i})}\frac{%
\partial P(y^{i})}{\partial y^{j}}\frac{\partial P(y^{i})}{\partial y^{k}}%
d\mu (y^{i})\right] \Delta y^{j}\Delta y^{k}
\end{equation}
The terms in square brackets are the elements of the Fisher information
matrix (while this is not the most general definition of the Fisher
information matrix, it is one that applies to the present case \cite
{Reginatto}. For the general case, see Ref. \cite{Kullback}). \ 

If $P$\ is defined over an $n$-dimensional manifold $M$ with (positive)
metric $g^{ik}$, there is a natural definition of the amount of information $%
I$ associated with $P$, which is obtained by contracting the metric $g^{ik}$
with the elements of the Fisher information matrix, 
\begin{equation}
I=\sum_{j,k}^{n}g^{ik}\int \frac{1}{P}\left( \frac{\partial P}{\partial y^{i}%
}\right) \left( \frac{\partial P}{\partial y^{k}}\right) d\mu (y^{i}).
\label{FIndim}
\end{equation}
In the case where $M$\ is the $n+1$ dimensional extended configuration space 
$QT$ (with coordinates $\{t,x^{1},...,x^{n}\}$) of a non-relativistic
particle of mass $m$, the natural metric is the one used to define the
kinematical line element in configuration space, which is of the form $%
g^{ik}=diag(0,1/m,...,1/m)$\ \cite{Synge}. \ Note that with this metric, it
is straightforward to generalize the results of the paper to the case of
diffusion in many space dimensions. \ In particular, we replace the velocity
field in eq. (\ref{sigma}) by the expression $v^{i}=g^{ik}\partial \sigma
/\partial x^{k}$, and the Lagrangian in eq. (\ref{LD}) by 
\begin{equation}
L_{D}=-\int \rho \left\{ \frac{\partial \sigma }{\partial t}+\frac{1}{2}%
\sum_{j,k}^{n}g^{ik}\left[ \frac{\partial \sigma }{\partial x^{i}}\frac{%
\partial \sigma }{\partial x^{k}}-\frac{D^{2}}{4}\frac{1}{\rho ^{2}}\frac{%
\partial \rho }{\partial x^{i}}\frac{\partial \rho }{\partial x^{k}}\right]
\right\} d^{n}xdt.
\end{equation}
In the case of one time and one space dimension, eq.(\ref{FIndim}) reduces
to eq. (\ref{FI}). \ 

To express $I$ in units of energy, we need to introduce a conversion factor
with units of action squared and multiply eq. (\ref{FI}) by this factor. \
In the case of the diffusion process, we can set the conversion factor
proportional to $D^{2}$, although it is also possible to introduce a
universal constant of action, such as $\hbar $, and set the conversion
factor proportional to $\hbar ^{2}$. \ 

\section{Appendix B}

We want to examine the extremum obtained from the fixed end-point variation
of the Lagrangian $L_{D}$. \ In particular, we wish to show the following:
given $\rho $\ and $\sigma $\ that satisfy eqs. (\ref{contsigma})\ and (\ref
{modH-J}), a small variation of the probability density $\rho
(x,t)\rightarrow \rho (x,t)^{\prime }=\rho (x,t)+\epsilon \delta \rho (x,t)$
for fixed $\sigma $ will lead to an increase in $L_{D}$, as well as an
increase in the Fisher information $I$.

We assume fixed end-point variations ($\delta \rho =0$ at the boundaries),
and variations $\epsilon \delta \rho $ that are well defined in the sense
that$\ \rho ^{\prime }$ will have the usual properties required of a
probability distribution (such as $\rho ^{\prime }>0$ and normalization).

Let $\rho \rightarrow \rho ^{\prime }=\rho +\epsilon \delta \rho $. \ Since $%
\rho $\ and $\sigma $\ are solutions of the variational problem, the terms
linear in $\epsilon $ vanish. If we keep terms up to order $\epsilon ^{2}$,
we find that

\begin{eqnarray}
\Delta L_{D} &\equiv &L_{D}(\rho ^{\prime },\sigma )-L_{D}(\rho ,\sigma ) 
\nonumber \\
&=&\epsilon ^{2}\frac{D^{2}}{8m}\int \int \left\{ \frac{(\delta \rho )^{2}}{%
\rho ^{3}}%
{\partial \rho  \overwithdelims() \partial x}%
^{2}-\frac{2\delta \rho }{\rho ^{2}}%
{\partial \rho  \overwithdelims() \partial x}%
{\partial \delta \rho  \overwithdelims() \partial x}%
+\frac{1}{\rho }%
{\partial \delta \rho  \overwithdelims() \partial x}%
^{2}\right\} dxdt+O\left( \epsilon ^{3}\right) .
\end{eqnarray}
Using the relation 
\begin{equation}
\frac{\partial }{\partial x}\left( \frac{\delta \rho }{\rho }\right) =\frac{1%
}{\rho }\frac{\partial \delta \rho }{\partial x}-\frac{1}{\rho ^{2}}\frac{%
\partial \rho }{\partial x}\delta \rho ,
\end{equation}
we can write $\Delta L_{D}$\ as 
\begin{equation}
\Delta L_{D}=\epsilon ^{2}\frac{D^{2}}{8m}\int \int \rho \left[ \frac{%
\partial }{\partial x}\left( \frac{\delta \rho }{\rho }\right) \right]
^{2}dxdt+O\left( \epsilon ^{3}\right) ,
\end{equation}
which shows that $\Delta L_{D}>0$ for small variations, and therefore the
extremum of $\Delta L_{D}$ is a minimum. \ Furthermore, since $\Delta
L_{D}\sim D^{2}$, it is the Fisher information term $I$\ in the Lagrangian $%
\Delta L_{D}$ that increases, and the extremum is also a minimum of the
Fisher information.

\end{document}